\documentclass[12pt,prl,preprint,english]{revtex4}

\usepackage{graphicx}
\usepackage{babel}

\providecommand{onlinecite}{\cite}

\makeatother

\begin{document}

\title{Superdense Crystal Packings of Ellipsoids}

\author{Aleksandar Donev$^{1,2}$, Frank H. Stillinger$^{4}$, P. M. Chaikin$^{2,3}$, and Salvatore
Torquato$^{1,2,4}$}

\thanks{S. T., A. D. and F. H. S. were supported in part by the Petroleum
Research Fund under Grant No. 36967-AC9, and by the National Science
Foundation under Grant Nos. DMR-0213706 and DMS-0312067. P. M. C.
was partially supported by NASA under Grant No. NAG3-1762.}

\address{$^{1}$Program in Applied and Computational Mathematics, Princeton
University, Princeton, NJ, 08544}

\address{$^{2}$Princeton Materials Institute, Princeton, NJ, 08544}

\address{$^{3}$Department of Physics, Princeton University, Princeton, NJ,
08544}

\address{$^{4}$Department of Chemistry, Princeton University, Princeton,
NJ, 08544}

\begin{abstract}
Particle packing problems  have fascinated people since the
dawn of civilization, and continue to intrigue
mathematicians and
scientists. Resurgent interest
has been spurred by the recent proof of Kepler's conjecture: the 
face-centered cubic lattice provides the densest packing of equal 
spheres with a packing fraction $\varphi\approx0.7405$ \cite{Kepler_Hales}.  
Here we report on the densest known packings of congruent ellipsoids. 
The family of new packings are crystal (periodic) arrangements of 
nearly spherically-shaped ellipsoids, and always surpass the densest lattice 
packing. A remarkable maximum
density of $\varphi\approx0.7707$ is achieved for both prolate and oblate 
ellipsoids with aspect ratios of $\sqrt{3}$ and $1/\sqrt{3}$,
respectively,
and each ellipsoid has 14 touching neighbors.
Present results do not exclude the possibility that even denser crystal
packings of ellipsoids could be found, and that a corresponding Kepler-like
conjecture could be formulated for ellipsoids.

\end{abstract}
\maketitle

\section{Background}

Dense packings of nonoverlapping particles have been employed
to understand the structure of a variety of many-particle
systems, including glasses \cite{Amorphous_solids}, crystals \cite{Aschcroft_Mermin}, 
heterogeneous materials \cite{Random_Materials}, and granular media \cite{Entropy_GM}.
The famous Kepler conjecture postulates that the densest packing of
spheres in three-dimensional Euclidean space has a packing fraction
(density) $\varphi=\pi/\sqrt{18}\approx0.7405$, as realized by stacking
variants of the face-centered cubic (FCC) lattice packing. It is
only recently that this conjecture has been proved \cite{Perfect_packing,Kepler_Hales}.
Very little is known about the most efficient packings of convex congruent
particles that do not tile three-dimensional space \cite{Conway_Packings,Zong_Packings}.
The only other known optimal three-dimensional result involves
infinitely long circular cylindrical particles: 
the maximal packing density $\phi_{\mbox{\scriptsize max}}=\pi/\sqrt{12}$ is attained by
arranging the cylinders
in parallel in the triangular lattice arrangement \cite{Densest_Ellipsoid}.
Of particular interest are dense 
packings of congruent ellipsoids (an affine deformation of a sphere)
with semi-axes $a$, $b$ and $c$ or, equivalently, with aspect ratios  $\alpha=b/a$ and $\beta=c/a$.

In two dimensions, it can easily be shown that the densest packing
of congruent ellipses has the same density as the densest packing of circles,
$\varphi=\pi/\sqrt{12}\approx0.9069$  \cite{Regular_figures,Combinatorial_Geometry}.
This maximal density is realized by an affine (linear) transformation
of the triangular lattice of circles. Such a transformation
leaves the density unchanged.
In three dimensions attempts at increasing the packing density
yield some interesting structures, at least for needle-like ellipsoids.
By inserting very elongated ellipsoids into cylindrical void channels
passing through the ellipsoidal analogs of the densest ordered sphere
packings (an affinely deformed face centered cubic or hexagonal close
packed lattice), congruent ellipsoid packings have been constructed whose density exceeds
$0.7405$ and approaches $0.7585$ in the limit of infinitely thin prolate
spheroids (ellipsoids of revolution), i.e., when $\beta=1$ and $\alpha\rightarrow\infty$
\cite{Densest_Ellipsoid,Wills_Densest}. 

However, there appears to
be a widespread belief that for nearly spherical ellipsoids the highest
packing fraction is realized by an affine transformation (stretch
by $\alpha$ and $\beta$ along two perpendicular axes) of the densest
sphere packing, preserving the density at $0.7405$. Mathematicians
have often focused on \emph{lattice packings}, where a single particle
is replicated periodically on a lattice to obtain a crystal packing.
For ellipsoids, a lattice packing is just an affine transformation
of a sphere packing, and therefore a theorem due to Gauss \cite{Conway_Packings,Zong_Packings} enables 
us to conclude that the densest lattice ellipsoid packing 
has $\varphi\approx0.7405$.  The next level of generality
involves nonlattice periodic packings (lattice packings with
a multiparticle basis), where a unit cell consisting of several ellipsoids
with at least two \emph{inequivalent} orientations is periodically replicated
on a lattice to fill Euclidean space. We will refer to these as \emph{crystal
packings}. 

We present here a family of crystal packings of ellipsoids that are denser
than the best lattice packing for a wide range of
aspect ratios in the vicinity of the sphere point $\alpha=\beta=1$, and for certain
aspect ratios yields the densest known ellipsoid packings with $\phi\approx 0.7707$.

\section{The Discovery}

We recently developed a molecular dynamics technique for generating
dense random packings of hard ellipsoids \cite{Event_Driven_HE}.
The simulation technique generalizes the Lubachevsky-Stillinger (LS)
sphere-packing algorithm \cite{LS_algorithm} to the case of ellipsoids.
Initially, small ellipsoids are randomly distributed and randomly oriented in
a box with periodic boundary conditions and without any overlap. The
ellipsoids are given velocities and their motion followed as they
collide elastically and also expand uniformly, while the unit cell
deforms to better accommodate the packing. After some time, a jammed
state with a diverging collision rate is reached and the density reaches
a maximal value.

Using this technique, we generated nonequilibrium random close packings
of ellipsoids and observed that for certain aspect ratios (close to
$\alpha\approx1.25$, $\beta\approx0.8$) the random packings had a density
as high as $0.735$, surprisingly close to what we believed was the
densest ordered packing (stretched FCC lattice) \cite{Jammed_MM}.
This brought into question what the maximal density really was for
those aspect ratios. Extensive experience with spheres has shown that
for reasonably large packings, sufficiently slowing down the growth
of the density, so that the hard-particle system remains close to
the equilibrium solid branch of the equation of state, leads to packings near the FCC lattice
\cite{Torquato_MRJ,Anu_order_metrics}.
This however requires impractically long simulation times 
for large ellipsoid packings. By running the simulation
for very small unit cells, from $4$ to $16$ particles per unit cell, we were able
to identify crystal packings significantly denser than the FCC lattice,
and subsequent analytical calculations suggested by the simulation results
led us to discover ellipsoid packings with a remarkably high density of $\phi \approx 0.7707$.
This result implies a lower bound on the maximal density of any packing of
congruent ellipsoids, namely, $\varphi_{\textrm{max}} \geq 0.7707$.

\section{Superdense Laminated Packings of Ellipsoids}

We now describe the construction of a family of superdense crystal packings
of ellipsoids. We start from the FCC lattice, viewed as a laminate
of face-centered planar layers of spheres, as illustrated in Fig.
\ref{FCC_layers}a. We similarly construct layers from the ellipsoids
by orienting the $c$ semiaxis perpendicular to the layer, while orienting
the $a$ and $b$ axes along the axes of the face-centered square
lattice defining the layer, as shown in Fig. \ref{FCC_layers}b. In
this process, we maintain the aspect ratio of the squares of side
$L$ of the face-centered square lattice defining the layer, i.e.,
we maintain
\begin{equation}
L=\frac{4\alpha}{\sqrt{1+\alpha^{2}}},
\end{equation}
which enables us to rotate the next layer by $\pi/2$ and fit it exactly
in the holes formed by the first layer. This two-layer lamination
is then continued \emph{ad infinitum} to fill all space. This  can be viewed
as a family of crystal packings with a unit cell containing two ellipsoids.

We can calculate the minimal distance $h$ between 
two successive layers (that preserves impenetrability) from the condition
that each ellipsoid touches four other ellipsoids 
in each of the layers above and below it. This gives a simple
system of equations (two quadratic equations and one quartic equation),
the solution of which determines the density to be
\begin{equation}
\phi=\frac{16 \pi \alpha \beta}{3 h L^2}.
\end{equation}
Notice that the axis perpendicular to
the layers can be scaled arbitrarily, without changing the density.
We can therefore just consider spheroids with $\beta=1$. The
density of this crystal packing as a function of the aspect ratio
$\alpha$ is shown in Fig. \ref{rho_alpha}a, and is higher than the
density of the FCC sphere packing for a wide range of aspect ratios
around the sphere point $\alpha=\beta=1$, symmetrical with respect to the
inversion of $\alpha$ between prolate and oblate ellipsoids (we consider
the prolate case in the equations in this section). Two sharp maxima
with density of about $0.770732$ are observed
when the ellipsoids in the face-centered layers touch six rather than
four inplane neighbors, as shown in Fig. \ref{rho_alpha}b,
i.e., when $L=2\alpha$,
which gives $\alpha=\sqrt{3}$ for the prolate and $\alpha=1/\sqrt{3}$
for the oblate case. These two densest-known packings of ellipsoids
are illustrated in the insets in Fig. \ref{rho_alpha}a, and in these
special packings each ellipsoid touches exactly $14$ neighboring
ellipsoids (compare this to $12$ for the FCC lattice).

\section{The Future}

There is nothing to suggest that the crystal packing we have presented
here is indeed the densest for any aspect ratio other than the trivial
case of spheres. Many other possibilities exist for laminated packings
with alternating orientations between layers, and one such example
is shown in Fig. \ref{Densest_layers}. More generally, we believe
it is important to identify the densest periodic packings of ellipsoids
with small numbers of ellipsoids per unit cell. This may be done using
modern global optimization techniques, as has been done for various
sphere and disk packing problems. However, this is a challenging project
due to the complexity of the nonlinear impenetrability constraints
between ellipsoids. In particular, the case of slightly aspherical
ellipsoids is very interesting, as the best packing will be a perturbation
of the FCC lattice with a broken symmetry, and should thus be easier
to identify. In Fig. \ref{rho_alpha}b we see that the density of our
crystal packing increases smoothly as asphericity is introduced, unlike
for random packings, where a cusp-like increase is observed near $\alpha=1$
\cite{Jammed_MM}. Is there a crystal packing which leads to a sharp
increase in density for slightly aspherical ellipsoids? Further multidisciplinary
investigations are needed to answer this and related questions. The
results of such investigations could be used to formulate a Kepler-like
conjecture for ellipsoids.

\bibliographystyle{unsrt}
\bibliography{References}

\begin{thebibliography}{10}

\bibitem{Kepler_Hales}
T.~C. Hales.
\newblock An overview of the kepler conjecture.
\newblock {\em LANL e-print Archive}, {\tt
  http://xxx.lanl.gov/math.MG/9811071}, 1998.

\bibitem{Amorphous_solids}
R.~Zallen.
\newblock {\em The Physics of Amorphous Solids}.
\newblock Wiley, New York, 1983.

\bibitem{Aschcroft_Mermin}
N.~W. Ashcroft, N.~D. Mermin, and D.~Mermin.
\newblock {\em Solid State Physics}.
\newblock International Thomson Publishing, 1976.

\bibitem{Random_Materials}
S.~Torquato.
\newblock {\em Random Heterogeneous Materials: Microstructure and Macroscopic
  Properties}.
\newblock Springer-Verlag, New York, 2002.

\bibitem{Entropy_GM}
S.~F. Edwards.
\newblock {\em Granular Matter (A. Mehta, editor)}, chapter The Role of Entropy
  in the Specification of a Powder, pages 121--140.
\newblock Springer-Verlag, New York, 1994.

\bibitem{Perfect_packing}
T.~Aste and D.~Weaire.
\newblock {\em The Pursuit of Perfect Packing}.
\newblock IOP Publishing, 2000.

\bibitem{Conway_Packings}
J.~H. Conway and N.~J.~A. Sloane.
\newblock {\em Sphere Packings, Lattices, and Groups}.
\newblock Springer-Verlag, New York, 3rd edition, 1999.

\bibitem{Zong_Packings}
C.~Zong.
\newblock {\em Sphere Packings}.
\newblock Springer-Verlag, New York, 1999.

\bibitem{Densest_Ellipsoid}
A.~Bezdek and W.~Kuperberg.
\newblock {\em Applied geometry and discrete mathematics: DIMACS Ser. Discrete
  Math. Theoret. Comput. Sci. 4 (P. Gritzmann and B. Sturmfels, editors)},
  chapter Packing Euclidean space with congruent cylinders and with congruent
  ellipsoids, pages 71--80.
\newblock Amer. Math. Soc., Providence, RI, 1991.

\bibitem{Regular_figures}
L.~\FToth.
\newblock {\em Regular figures}.
\newblock Pergamon Press, 1964.

\bibitem{Combinatorial_Geometry}
J.~Pach and P.~K. Agarwal.
\newblock {\em Combinatorial Geometry}.
\newblock Wiley-Interscience, 1st edition, 1995.

\bibitem{Wills_Densest}
J.~M. Wills.
\newblock An ellipsoid packing in $e^3$ of unexpected high density.
\newblock {\em Mathematika}, 38:318--320, 1991.

\bibitem{Event_Driven_HE}
A.~Donev, S.~Torquato, and F.~H. Stillinger.
\newblock An event-driven molecular dynamics algorithm for nonspherical
  particles: Ellipses and ellipsoids.
\newblock In preparation, 2003.

\bibitem{LS_algorithm}
B.~D. Lubachevsky and F.~H. Stillinger.
\newblock Geometric properties of random disk packings.
\newblock {\em J. Stat. Phys.}, 60:561--583, 1990.
\newblock See also Ref. \onlinecite{LS_algorithm_3D}.

\bibitem{Jammed_MM}
A.~Donev, I.~Cisse, D.~Sachs, E.~A. Variano, F.~H. Stillinger, R.~Connelly,
  S.~Torquato, and P.~M. Chaikin.
\newblock Jammed ellipsoids beat jammed spheres.
\newblock Submitted for publication, 2003.

\bibitem{Anu_order_metrics}
A.~R. Kansal, S.~Torquato, and F.~H. Stillinger.
\newblock Diversity of order and densities in jammed hard-particle packings.
\newblock {\em Phys. Rev. E}, 66:041109, 2002.

\bibitem{Torquato_MRJ}
S.~Torquato, T.~M. Truskett, and P.~G. Debenedetti.
\newblock Is random close packing of spheres well defined?
\newblock {\em Phys. Rev. Lett.}, 84:2064--2067, 2000.

\bibitem{LS_algorithm_3D}
B.~D. Lubachevsky, F.~H. Stillinger, and E.~N. Pinson.
\newblock Disks vs. spheres: Contrasting properties of random packings.
\newblock {\em J. Stat. Phys.}, 64:501--525, 1991.
\newblock Second part of Ref. \onlinecite{LS_algorithm}.

\end{thebibliography}

\newpage

\section{Figures}

\begin{figure}[!h]

\caption{\label{FCC_layers} \emph{Part a (top):} The face-centered cubic
packing of spheres, viewed as a laminate of face-centered layers (bottom
layer is colored purple and the top layer yellow). \emph{Part b (bottom):}
A nonlattice layered packing of ellipsoids based on the FCC packing
of spheres, but with a higher packing fraction.}
\end{figure}

\begin{figure}[!h]

\caption{\label{rho_alpha}
\emph{Part a (top):}
The density of the laminate crystal packing of
ellipsoids as a function of
the aspect ratio $\alpha$ ($\beta=1$). The point $\alpha=1$ corresponding
to the FCC lattice sphere packing is shown, along with the two sharp
maxima in the density for prolate ellipsoids with $\alpha=\sqrt{3}$
and oblate ellipsoids with $\alpha=1/\sqrt{3}$, as illustrated in
the insets.
\emph{Part b (bottom):} The layers of the densest
known packing of ellipsoids, as illustrated in part a. 
The layers can be viewed as either face-centered
or triangular. 
}
\end{figure}

\begin{figure}[!h]

\caption{\label{Densest_layers}
Another computer-generated
layered packing of prolate ellipsoids ($\alpha=\sqrt{3}$) with a zig-zag-like variation in the
orientations of the ellipsoids in adjacent layers. This packing is
slightly denser than the FCC lattice ($\varphi=0.7411$), but we did not attempt to find
the optimal configuration.}
\end{figure}

\setcounter{figure}{0}

\newpage
\begin{figure}[!h]
\begin{center}\includegraphics[width=0.40\paperwidth,keepaspectratio]{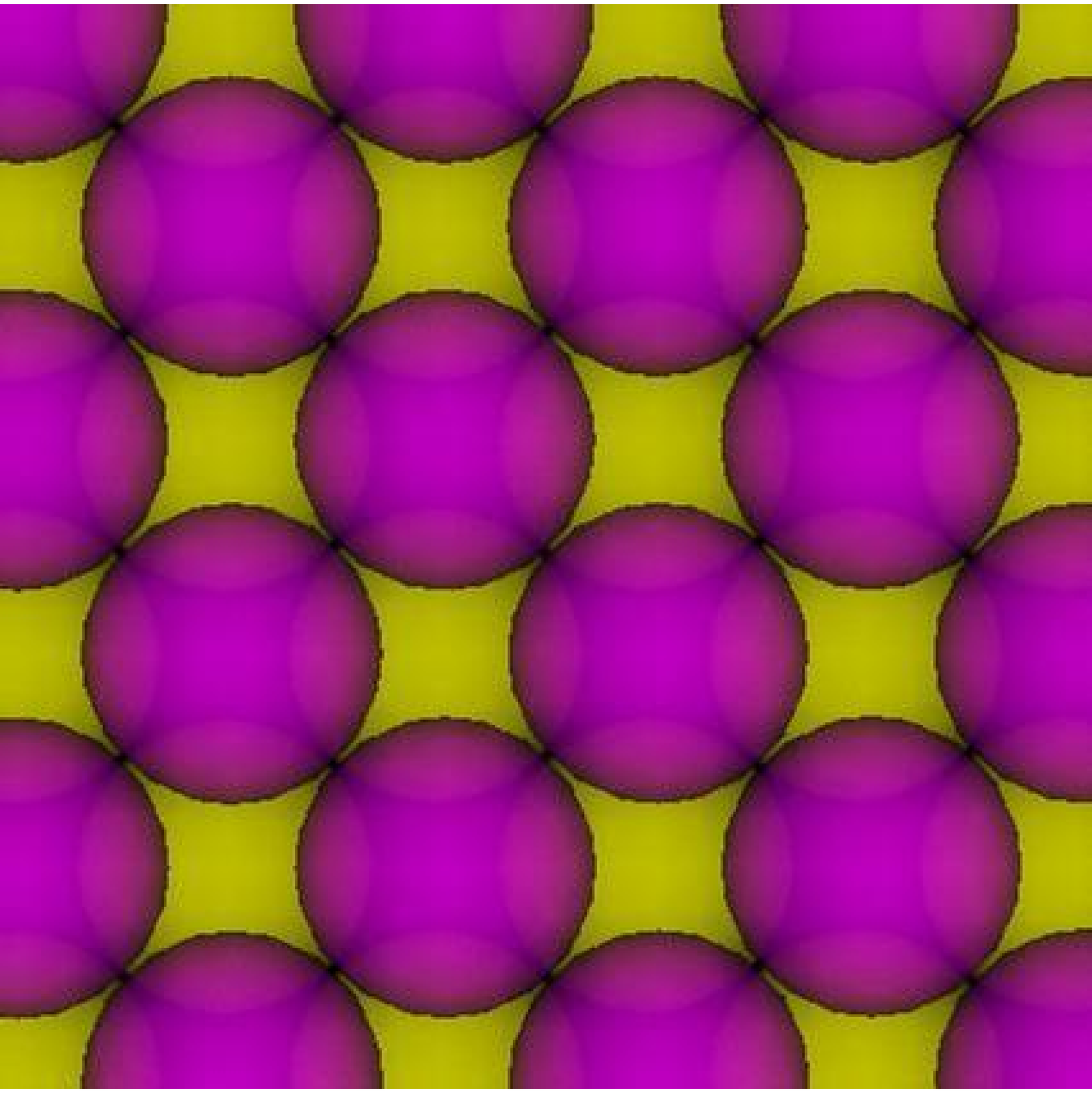}\end{center}
\begin{center}\includegraphics[width=0.40\paperwidth,keepaspectratio]{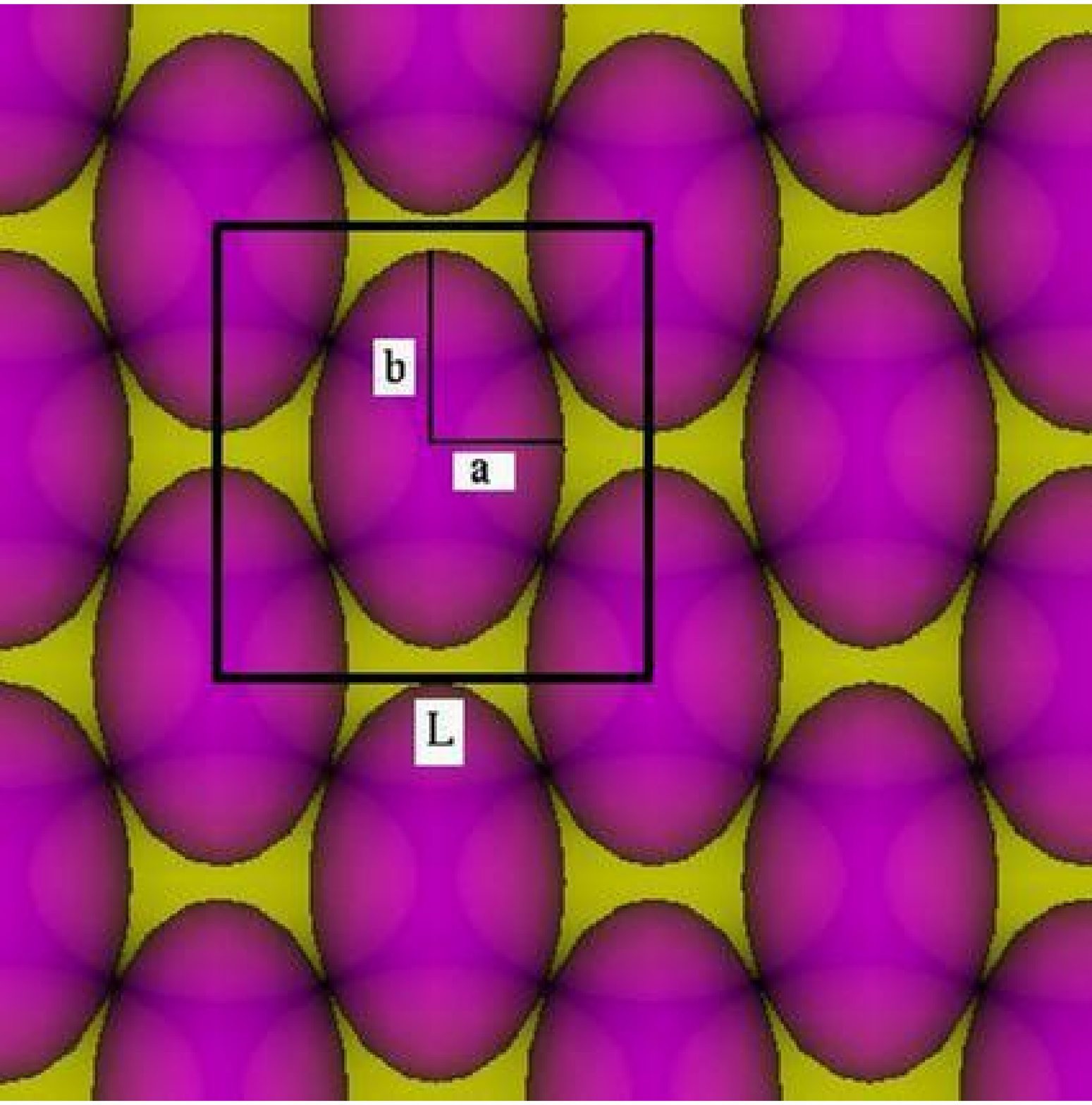}\end{center}

\caption{\emph{Donev et al.}}
\end{figure}

\newpage
\begin{figure}[!h]
\begin{center}\includegraphics[height=0.50\paperwidth,keepaspectratio]{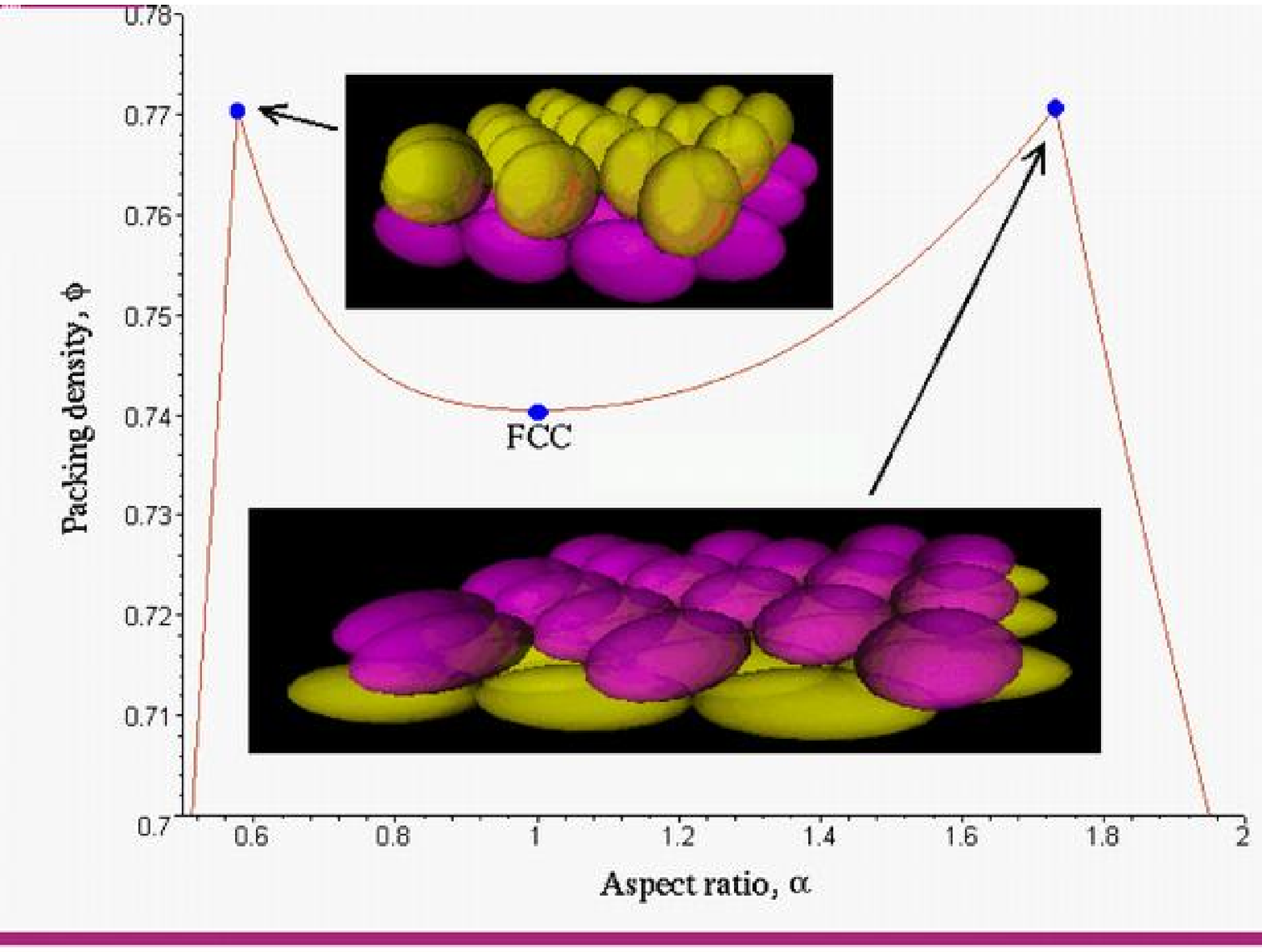}\end{center}

\begin{center}\includegraphics[height=0.30\paperwidth,keepaspectratio]{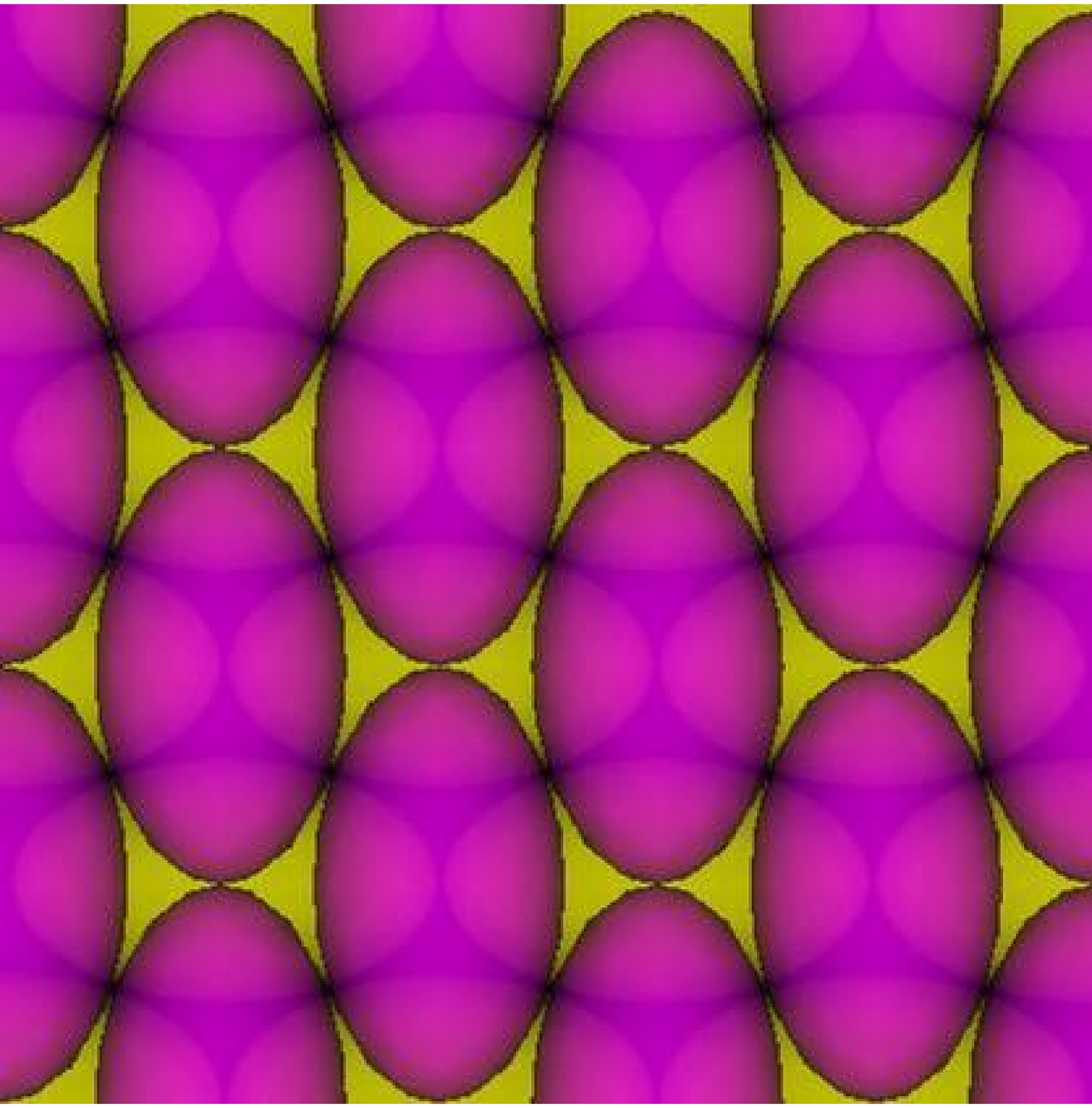}\end{center}

\caption{\emph{Donev et al.}}
\end{figure}

\newpage
\begin{figure}[!h]
\begin{center}\includegraphics[width=0.40\paperwidth,keepaspectratio]{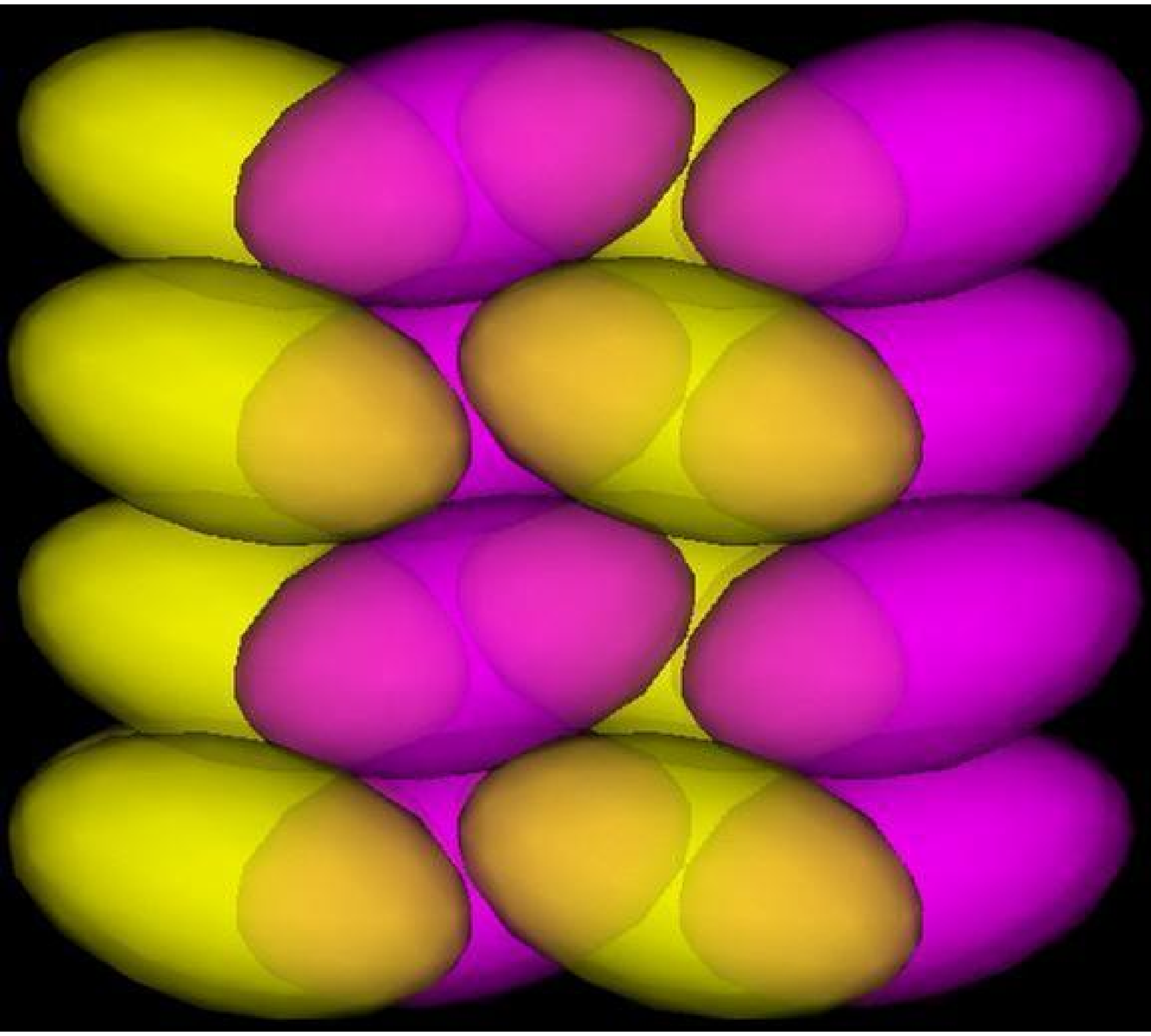}\end{center}

\caption{\emph{Donev et al.}}
\end{figure}

\end{document}